\documentstyle[12pt,aasms]{article}
\tighten
\eqsecnum

\slugcomment{ draft \today}

\begin{document}

\font\twelvei = cmmi10 scaled\magstep1 
       \font\teni = cmmi10 \font\seveni = cmmi7
\font\mbf = cmmib10 scaled\magstep1
       \font\mbfs = cmmib10 \font\mbfss = cmmib10 scaled 833
\font\msybf = cmbsy10 scaled\magstep1
       \font\msybfs = cmbsy10 \font\msybfss = cmbsy10 scaled 833
\textfont1 = \twelvei
       \scriptfont1 = \twelvei \scriptscriptfont1 = \teni
       \def\mit{\fam1 }
\textfont9 = \mbf
       \scriptfont9 = \mbfs \scriptscriptfont9 = \mbfss
       \def\bmit{\fam9 }
\textfont10 = \msybf
       \scriptfont10 = \msybfs \scriptscriptfont10 = \msybfss
       \def\bmsy{\fam10 }

\def\etal{{\it et al.~}}
\def\eg{{\it e.g.}}
\def\ie{{\it i.e.}}
\def\lsim{\raise0.3ex\hbox{$<$}\kern-0.75em{\lower0.65ex\hbox{$\sim$}}} 
\def\gsim{\raise0.3ex\hbox{$>$}\kern-0.75em{\lower0.65ex\hbox{$\sim$}}}
\def \msol {\rm{M}$_\odot$}
\def \msol {\rm{M}$_\odot$}
\def \mdot {\rm{M}$_\odot$~yr$^{-1}$}
\def \lam {$\lambda$}
\def \kms{km~$\rm{s}^{-1}$}
\def \cc{$\rm{cm}^{-3}$}
\def \arcs{\char'175}
\def \lam{$\lambda$}
\def \micra{$\mu$m}

\title{Precessing Jets and Molecular Outflows: A 3-D Numerical Study}

\author{J. A. Cliffe \altaffilmark{1}, Adam Frank \altaffilmark{2} 
and T.W. Jones \altaffilmark{3}}
\affil{Department of Astronomy, University of Minnesota,
    Minneapolis, MN 55455}
\altaffiltext{1}{e-mail: cliffe@astro.spa.umn.edu}

%\altaffiltext{1}{ Department of Astronomy, University of Minnesota,
\altaffiltext{2}{Hubble Fellow, e-mail: afrank@astro.spa.umn.edu}
\altaffiltext{3}{e-mail: twj@astro.spa.umn.edu}
\clearpage

%\begin{abstract}

We present 3-D numerical hydrodynamical simulations of precessing 
supersonic heavy jets to 
explore their evolution, how they differ from straight jets and how well
they serve as a model for generating molecular outflows from Young Stellar Objects. 
The dynamics are studied with a number of high resolution 
simulations on a Cartesian grid (128x128x128 zones) using a high order
finite difference method. A range of cone angles and precession rates
were included in the study. 
Two higher resolution runs (256x256x256 zones) were made for comparison
in order to confirm numerical convergence of global flow characteristics.
Morphological, kinematical and dynamical characteristics of precessing 
jets are described and compared to important properties of straight jets and 
also to observations of YSOs. 

In order to examine the robustness of precessing jets as a mean to 
produce molecular outflows around Young Stellar Objects, ``synthetic 
observations'' of the momentum distributions of the simulated precessing
jets are compared to observations of molecular outflows. 
It is found that precessing jets match better the morphology,
highly forward driven momentum and momentum distributions along the
long axis of molecular outflows than do wind-driven or straight 
jet-driven flow models. 


\section {Introduction}

There is now overwhelming evidence that stars of low-to-intermediate
mass experience 'vigorous' episodes of mass loss during their
evolution to the main sequence.  Observations of Young Stellar Objects
(YSOs hereafter) have revealed these ``outflows" in a
variety of forms, including strong stellar winds, rapidly moving H-H
objects, high velocity maser sources, and shock-excited molecular
hydrogen emission regions.  The high-velocity molecular outflows and
well-collimated optically visible jets are two particularly striking 
and ubiquitous
forms of YSOs outflow. Kinetic energies
derived for the out-flowing gas are on the order of $10^{43} -
10^{47} $ ergs, representing a significant energy input.  Such a large
energy budget has important implications for the study of YSOs since these
outflows appear to be intrinsic to the star formation process, as well
as producing important effects on the molecular clouds where stars are
born (Lada 1985).

Understanding the outflows is crucial to understanding the origin of
stars. A critical issue involves attempts to unify the
apparently disparate phenomena of YSO outflows into a common theoretical
paradigm. For example, it has been observed that molecular outflows, jets and H-H
objects are all sometimes associated with the same YSO (Masson \&
Chernin 1993).  However, while it is agreed that these phenomena are
spatially adjacent, it is not clear whether they are related causally.
The possible link between jets and molecular outflows has recently been the 
subject of considerable study. The question
can be phrased: Are jets the driving sources for the molecular
outflows, or are jets and molecular outflows the result of
intrinsically different kinds of phenomena (such as different kinds of
winds) associated with the central YSO?  Models that rely on winds for
producing the molecular outflows have been shown to produce line
profiles with the wrong shape and require unrealistic physical
conditions (Masson \& Chernin 1992).  Jet-driven models, another 
possible generation mechanism, come in two
flavors. Either the momentum is imparted to the ambient medium
impulsively through the bow shock (Chernin {\it et al.} 1994) 
or continuously 
thorough
entrainment in a turbulent boundary layer (Stahler 1994).  While both
forms of {\it straight} jet-driven models are more successful than
wind-driven models in explaining some aspects of outflows, each
fails to recover the full suite of outflow characteristics.

In this paper we address the relation between YSO jets and molecular
outflows by focusing on the places in the database where the
conventional jet-driven models fail.  Using fully three-dimensional
numerical simulations we explore a model in which the outflows are
driven by {\it precessing} YSO jets interacting with the surrounding
dense molecular cloud.  Although such a scenario has been suggested in
the literature (see references below), except for a preliminary report
by us (Cliffe {\it et al.} 1995), rigorous modeling of the full
three-dimensional evolution of the resulting flow pattern is lacking
(see Biro, Raga, \& Cant\'o 1995 for an excellent 2-D treatment of the
problem).  With some notable exceptions involving analytic studies
(e.g., Raga, Cant\'o \& Biro 1993a) the basic jet physics of
precessing jets remains essentially unexplored.  Thus, while we are
ultimately interested in the relationship between jets and molecular
outflows, our simulations will also cover new territory in the
propagation of YSO jets. These structures are fundamentally
three-dimensional and complex.  Thus less complete methods are likely
to be inadequate.  Three-dimensional simulations are, however, computationally
expensive.  This is also relatively unexplored territory. So, we
concentrate in this paper on the idealized case of a jet propagating
into a constant density medium and where the interaction is governed
by a polytropic equation of state rather than including a realistic
cooling function.  We wish to focus on the so-called bow shock models
of jet-driven outflow (Masson \& Chernin 1993).  Thus, our goals in
the present study are to understand the basic physics of precessing
jets, to examine the morphological and kinematical differences between
precessing and straight jets, and, finally, to compare precessing and
straight jets' abilities to produce certain key observed
characteristics of molecular outflows.

\section{Background} 

Since the discovery of molecular outflows associated with YSOs, it has
generally been believed that some jet and/or wind combination with an
array of ionized and neutral components is responsible for driving the
flows.  Models include momentum-conserving shells driven by a wide
angle wind (Shu {\it et al.} 1991; Cabrit 1992; Masson \& Chernin 1992), a
simple, straight jet/bow shock driven model (Masson \& Chernin 1993)
and jets with a viscous boundary layer (Raga {\it et al.} 1993b; Stahler
1993; 1994). Thus far all of these scenarios have had only moderate
success in accounting for observed flow characteristics (Masson \&
Chernin 1993).  There are several common, important observed
properties that current (stellar-wind or straight jet-driven)
scenarios have been unable to explain satisfactorily.  In this paper
we will focus on three particularly troublesome properties (Masson \&
Chernin 1993); namely:

1) The degree of collimation varies among different objects. Some 
outflows have been observed to be highly collimated,
with length to width ratios of as great as 20:1. Others are very wide, with
this ratio near unity.

2) The momentum and mass distributions along the axis in the outflows tend to
peak near the middle of each lobe with minima near both
the star and the end of the lobe.  This result was found in a
recent study by Chernin \& Masson (1995), who averaged across
cuts perpendicular to the flow axis for six outflows to produce profiles of
mean momentum
($d{\bar P}/dz$) and mass per unit length ($d{\bar m}/dz$). Here
${\bar P} = P/w$ and ${\bar m} = m/w$ where $w$ is the width of the lobe
at a distance $z$ from the star.
In addition
to finding peaks in the middle of each lobe, Chernin and Masson
concluded that the underlying velocity fields in their
sample were relatively uniform along the length of each outflow.

3) The momentum in the outflows is primarily forward driven. ``Forward
driven'' means most of the velocity vectors in the flow are oriented
along the long axis of the lobes.  This kind of flow pattern would be
difficult to achieve if the molecular outflows formed as ``energy
conserving'' wind blown bubbles (Masson \& Chernin 1993).  In such a
case the lobes would be inflated by the pressure of shocked stellar
wind material.  Since thermal pressure always acts normal to the
surface of the lobe one would be forced to predict significant
velocities transverse to the axis of the lobe.  Observationally these
transverse motions would appear as both red and blue shifted velocity
components from {\it each} lobe of a bipolar outflow.  Studies of
molecular outflows, however, have not revealed the presence of
transverse motions. In general, blue (red) shifted material dominates
in the blue (red) shifted bipolar lobe. A recent study by Lada \& Fich
(1995) emphasized this point as their observations of NGC 2064G reveal
20:1 ratios of blue to red-shifted gas in the blue lobe.

Masson and Chernin (1993, 1995) have suggested that wandering or
precessing jets may produce flows fitting the observational
constraints better.  In the wandering jet model the jet head drives
different parts of the ambient cloud as it changes direction. This
solves many of the problems of other models.  First a wider lobe is
produced.  Second, and more importantly, momentum from the flow can be
transferred to the ambient medium at more than one location and mostly in
the "forward" direction of the jet itself. Chernin
and Masson (1995) also argue that such wandering jets will 
match better their momentum distribution observations.  However, their
arguments were heuristic, and they emphasized the need for accurate
numerical calculations.  In their work on NGC 2064G Lada \& Fich
(1995) also suggest that wandering jets may account better for the
large forward driven velocities in molecular outflows. Despite these
several suggestions in the literature, there are no fully non-linear,
time-dependent, and three-dimensional calculations of the resulting
flow patterns to serve as tests of the ideas they represent.

Beyond the above, indirect indications, there is a growing body of
direct evidence for HH jet precession in YSO outflows.  Precessing
jets seem to be observed in the HH 80/81 system (Marti, Rodriguez, \&
Reipurth 1993), in Serpens (Curiel et al. 1993), in HH 7-11 (Lightfoot
\& Glencross 1986), and in the chain of HH 34 objects (Bally \& Devine
1994).

Although direct evidence for ``naked'' wandering jets exists, observing
wandering jets inside molecular outflows is a more difficult task.
Still, some evidence for wandering jets inside of molecular outflows
can be seen, for example in the curved structure of the VLA 1623
molecular outflow (Dent, Matthews, \& Walther 1995).  The
observational situation appears to be improving, however, as Gueth \&
Guilloteau 1995 have recently provided strong evidence for a
precessing jet inside the L1157 molecular outflow.  In addition,
Plambeck and Snell (1995) have shown the L1551 outflow shell to have a
clumpy structure whose configuration is suggestive of an internal
precessing or wandering jet.

\section{The Numerical Model and Methods}

We have carried out several numerical experiments with fully 3-D
gasdynamical methods to address some of the above issues.
As a simple model for wandering jets that should contain many of the
essential features of such flows we 
simulated jets precessing at a steady rate, $\Omega = 2\pi/\tau_p$,
around the $z$-axis at a constant cone half angle, $\theta$.
The parameters $\Omega$
and $\tau_p$ will be termed the precession rate and precession period,
respectively. For the present simulations we assume the jet enters the
volume with fixed speed, $v_o$, and density, $\rho_j = \chi
\rho_e$, where $\rho_e$ is a uniform external gas density and $\rho_e$
is the jet density. Thus, the
Cartesian velocity of the jet as it is injected into the 3-D computational
space (at the origin ($x = y = z = 0$) satisfies:
\begin{equation}
v_x = v_o \sin{\theta} \cos{\phi}
\label{jetvel1}
\end{equation}
\begin{equation}
v_y = v_o \sin{\theta} \sin{\phi}
\label{jetvel2}
\end{equation}
\begin{equation}
v_z = v_o \cos{\theta}
\label{jetvel3}
\end{equation}
where $\phi = \Omega \times t + \phi_o$. For the simulations reported
here we set $\phi_o = 0$. We also assume that the injected jet material is
in pressure equilibrium with the external medium, so that the
jet sound speed, $c_j = \sqrt{\gamma p_o/\rho_j}$, is related to the 
external sound speed, $c_e$, as $c_j = c_e/\sqrt{\chi}$. Thus, the
Mach number of the jet with respect to the external medium is, $M_j = v_o/c_e$.

If the motion were purely kinematic, the locus of the jet beam would form a
conical helix with constant pitch, $\delta z = 2\pi v_o
\cos{\theta}/\Omega$.  Projected onto the $x-y$ plane, the trajectory
would be an Archimedian spiral of instantaneous projected radius, $R_s = v_o t
\sin{\theta}$. It is important to recognize, however, that the
actual motion of a fluid element in this idealized kinematic jet is
purely radial in three-space from the point of injection; that is, $v_{\phi} =
0$. Interaction with the ambient medium will substantially modify this
motion on a dynamical timescale that can be characterized in units,
$\tau_d = \sqrt{\chi} a/v_o$, where $a$ is the initial radius of the
inflowing jet (see \S 4). 
Henceforth we will use ``natural'' units for time length
and density;  namely, $t' = t/\tau_d$, $l' = l/a$, $\rho' = \rho/\rho_e$.
In these units the jet velocity is $v_o' = \sqrt{\chi}$ and the
two characteristic sound speeds are $c_e' = \sqrt{\chi}/M_j$ and
$c_j' = 1/M_j$. The pitch angle becomes $\delta z' = 2\pi \sqrt{\chi}
\cos{\theta}/\Omega'$, where $\Omega' = \Omega \tau_d$.  For simplicity we 
will drop the primes in our further discussions.

There are a minimum four free parameters needed to describe the jets;
namely, $\chi$, $M_j$, $\theta$ and $\tau_p$.  Physically, we expect
from studies of straight, cylindrical jets that the density contrast
between the jet and the external medium, $\chi$, and the Mach number
of the jet (e.g., Norman {\it et al.}, 1982; Blondin {\it et al.}
1990; Chernin {\it et al.} 1994, Bodo {\it et al.} 1995) will have
considerable influence on behaviors. There are qualitative
distinctions between ``light'' jets ($\chi < 1$) and ``heavy jets''
($\chi >1$), as well as between subsonic and supersonic jets.  Stellar
jets appear to be in the ``heavy jet'' and supersonic regimes (Mundt
{\it et al.} 1987; Morse {\it et al.} 1992, 1993), so we shall focus
our attentions there.  These parameters are not tightly constrained by
observation, but values for $\chi$ between 10 and 100 and $M_j \approx
10$ are often quoted (Hartigan, Morse, \& Raymond 1994; Stone \&
Norman 1993).  For simplicity we consider a single jet Mach number,
$M_j = 10$, as being representative. Appropriate values for the other
parameters need to be established in concert. We are particularly
concerned about precession with a wide enough cone to generate the
``global'' bow shock identified by Cliffe {\it et al.} (1995). As we shall see,
it is also important to consider that once material in the most
forward regions of a precessing jet, the jet ``head'', begins to
decelerate after a time $t \sim 1$ (in our natural units), it may
merge with younger, trailing jet elements, if their trajectories are
close enough.  Thus, to be interesting, our simulations need to extend
over a time $t > 1$ and must contain more than a single precession
period for the jet. The first condition requires a grid that extends to a
vertical height, $z_{max} > \sqrt{\chi}\cos{\theta}$.  The second
condition requires $z_{max}/ \delta_z > 1$ or a precession period
$\tau_p < 1$.  The grid must
also extend far enough in the $x$ and $y$ directions to enclose both
the wandering jet and its global bow shock.  We want to simulate jets
having sufficient wobble that material ejected to opposite sides of
the cone do not have overlapping trajectories without
deflection. Otherwise, the concept of a ``global'' bow shock has no
real meaning. If $\sin{\theta} > 1/\sqrt{\chi}$ the wobble is sufficient
that the trajectories part before $t=1$.
 We chose two precession angles, $\theta = 12^{\rm o}$ and
$26^{\rm o}$ as representative and practical to compute.  A cubical
grid with $z_{max} = 128$ along with $\chi = 80$ enables us to
enclose the jet long enough to study interesting dynamical
developments. The nominal times for these jets to cross to the top of
the grid would be $t_{cross} = 1.25~(\theta = 26^o)$ and $t_{cross} =
1.15~(\theta = 12^o)$.  In order to satisfy the condition concerning
multiple turns in the jet we simulated at each precession angle jets
with periods $\tau_p = (1/2) t_{\rm cross}$ and $\tau_p = (1/5)
t_{\rm cross}$.  For a control model we also computed a straight
steady, jet that was identical in every respect to the precessing ones
except for the precession itself.

Radiative cooling effects are likely to be important in determining
the detailed flows associated with real stellar jets. However, for
these preliminary explorations we can capture much more economically
the enhanced compressibility of strongly radiative flows by modeling
with an isothermal, polytropic equation of state.  Here, therefore, we
assume a gas equation of state, $p \propto \rho^{\gamma}$. Most of our
simulations were carried out using $\gamma = 5/3$, but we also carried
out one run with $\gamma = 1.1$, which we shall term ``isothermal''.

Our simulations were carried out using a fully 3-Dimensional (3-D)
gasdynamics code based on a Total Variation Diminishing (TVD) scheme
(see Ryu {\it et al.} 1993 for details). The scheme is a conservative,
explicit second-order accurate finite difference method that uses a
Roe-type Riemann solver to estimate upwind fluxes. Our
implementation is Eulerian. For meaningful results it is important
that we fully resolve shocks and contact surfaces that form within the
jet. Since TVD codes such as the one we are using
spread shocks over 2-3 zones and contact surfaces
over as many as 5 zones, it is clear that the jet must span a
significantly greater set in order to be captured dynamically. In
addition, multi-dimensional disordered flows that might develop are
subject to substantial numerical diffusion on scales less than about
10 zones (Ryu \& Goodman 1994).  Thus, to allow possible jet surface
shear instabilities to form, we set ten zones as a lower bound for the
jet radius, $N_a$. For most of our experiments we did use $N_a = 10$,
but to confirm these results we repeated two of them using twice that
resolution, $N_a = 20$. In fact, we found no significant differences
in the behaviors between the comparable runs, except for sharper
definition of the structures as described above and as normally
expected in such simulations. To meet the earlier constraints imposed
on the simulations our $N_a = 10$ simulations were carried out on a
$128^3$ zone grid, while the $N_a = 20$ simulations used a $256^3$
grid.  Model properties are summarized in Table 1.

The simulations are initiated with the jet pre-formed inside ghost
zones under the center of the cube bottom ($z = 0$).  Within a
circular cylinder of radius $N_a$ the velocity is defined at each time
step according to equations [3.1-3.3] The density and pressure in this
region are maintained at $\rho_j = \chi$ and $p_e = \chi /(\gamma
M^2_j)$, respectively. To avoid a numerical blending between the jet
and surrounding material at its origin resulting from the oblique jet
velocities, we placed a stationary three zone wide ``collar'' around
the jet within the ghost zones . In the collar $\rho = \rho_e = 1$ and
$\vec v = 0$.  Except for the jet and the collar, all boundaries are open
or continious. Simulations end when the bow shock of the jet
passes through the top of the cube.

\section{Results}

Our objectives for this study are to understand better the formation
and character of the global bow shock reported for precessing jets by
Cliffe {\it et al.} (1995), to begin examination of the basic dynamics
of precessing jet material as well as issues associated with
entrainment of ambient material into the outflow. Since the stimuli
for these calculations were observational, we also need to make some
qualitative comparisons with the properties seen in real molecular
outflows. In this section we will outline the salient morphological,
kinematical and dynamical features exposed by our simulations. The
following section will look at some kinematical issues that relate to
observations of molecular outflows.

\subsection{Morphology and Jet Dynamics}

Three figures can serve to illustrate most of the prominent
morphological features seen in all the simulations.  Figure 1 presents
volume rendered density images for models 1,2,3 and 4 for $t \approx
1.1-1.2$. Figure 2 shows the density distribution in the $x-z$ plane
($y = 0$) of model 7 at four selected times, while Figure 3
illustrates distributions of both the density and pressure for model 7
at $t \approx 1.2 $. To start, we simply describe the flows, but will
follow with some discussion of the physics of the associated dynamics.
All of these images represent cases with $\gamma = 5/3$. We will
contrast isothermal flows later.  From Figure 1 we can see the conical
helix form of the jets enclosed by the global bow shocks.  Figure 2
demonstrates that the jet material becomes strongly compressed in
roughly the radial direction from the jet origin. It shows that over
time the leading jet material is strongly decelerated and eventually
experiences a rear collision from material in the next turn of the
helix, so that the two turns merge. The figure also makes apparent a
wake behind the first turn in the helix that is reinforced by
following turns. So, much of the volume inside the global bow shock,
sometimes called the ``jet shroud'', is strongly evacuated. Figure 3
shows that there is a broad, relatively uniform high pressure region
between the bow shock and the lead turn in the jet. That uniform
pressure extends within the lead turn, in fact, ending at strong
shocks penetrating through the jet material. Those ``jet'' shocks
account for both the jet compression and the deceleration of the head
of the jet.  Also visible in Figures 2 and 3 are weaker, secondary bow
and jet shocks formed by the motions of younger, following jet turns
within the confines of the global bow shock. These were predicted by
Raga, Cant\'o \& Biro (1993).

It is informative to contrast the characteristics just mentioned with
those of more traditional straight, steady jets. It should be kept in
mind that these are ``heavy'' jets, so that some details are
automatically different from the more commonly studied ``light''
jets. Our observations are based on simulations we carried out using
methods identical to those we employed for the precessing jets, but
they are also apparent from previously published results (Norman {\it
et al.} 1982). Straight jets also are surrounded by a bow shock, of
course. Those associated with the precessing jets differ primarily in
the greater breadth coming out of the precession. There is also a
reverse, jet or ``terminal'' shock near the head of the straight jet,
analogous to the leading jet shock seen in Figures 2 and 3.  In the
astrophysical jet literature the bow and jet shocks are often called
``working surfaces'', since they are regions of sharp energy
dissipation and likely sites for enhanced emission.  The gas pressure
between the bow shock and the jet shock is roughly constant, just as
for precessing jets.  For a straight jet, new material is continuously
``colliding'' with material decelerated by the jet shock, so that the
head of the jet does not experience a continuous deceleration like
those of precessing jets must. Instead, the dense, shocked head of a
straight jet simply gets longer with time. In effect, the straight jet
is continuously experiencing rear collisions that add new momentum and
mass to the flow of the head.  Although the mean circum-jet density is
somewhat lowered from the ambient gaseous medium do to the lateral
expansion of the bow shock, there is no strongly evacuated cavity,
such as we see in the precessing jets. Note that heavy straight jets
are not expected to be surrounded by a cocoon of shocked jet material,
in contrast to light jets.  The interia of the jet material is too
great and the sound speed of the shocked jet material too low for a
strong back flow to form. That feature also carries over to the
precessing jets we have simulated.  But, in contrast to straight jets,
different sections of the precessing jets evolve almost independently,
except for relatively gentle influence from wakes and possible
collisions with adjacent turns of the jet if one has been
decelerated. These properties make precessing jets rather similar in
significant ways to ``restarting jets'' (Clarke \& Burns 1991) and
supersonic ``bullets'' (e.g., Jones, Kang \& Tregillis 1994).

For both straight and precessing jets, dynamics of the jet material
can be described in terms of the influence of the reverse, jet
shock. It is simple in either case to show (e.g., Jones \& Kang 1994)
that the speed of that shock through the jet material, $v_{js} \approx
1$. Since the sound speed within the jet,
$c_j = 1/M_j$, the Mach number of the jet shock $\sim M_j$.  In
precessing jets each turn of the jet forms its own set of ``bow'' and
''jet'' shocks. However, it is apparent from Figures 2 and 3, because
the densities and pressures encountered by following turns of the jet
are low, that those shocks are much weaker than the ones associated
with the head of the jet. Thus, we will ignore them in our discussion
of jet dynamics. It is possible on the other hand that these shocks
could be effective ``working surfaces'' in the traditional sense that
they are regions where enhanced emission might be generated. As
discussed below, they are responsible for accelerating ambient gas
within the shroud and help to explain the distinctive kinematics for
some YSO outflows. In particular, Plambeck \& Snell (1995) have
recently studied two bright ``high-velocity'' CS emission regions which
appear on opposite sites of the central star in the L1551 outflow.  
Their results show that these clumps can be
interpreted as low velocity shocks propagating into the lobes from
within.  While they propose an uncollimated wind as the source of the
shock waves, the weak secondary shocks seen in our simulations provide
an alternative and, perhaps more attractive, hypothesis.  The flow
pattern Plambeck \& Snell (1995) observe, showing red and
blue-shifted motion on opposite sides of the star, would occur
naturally as the result of the weak secondary shocks and the
point-symmetry inherent to a two sided precessing jet.

The sound crossing time, $t_{sc} \sim 1/c_j \sim M_j$, within
unshocked regions of the jets is very long compared
to the duration of our simulations. Coupled with the fact that jet
material only begins to be strongly decelerated on a timescale $t \sim
1$, this means that we should not expect to see much evidence of
disruptive instabilities within the jet (e.g., Bodo {\it et al.} 1995,
Jones, Kang \& Tregillis 1994), unless we follow them over an interval
several times greater than we have done.  In fact, we find no evidence
for disruptive instabilities in any of our simulations reported here.

At several points we have alluded to the timescale for deceleration of
jet material  being $t \sim  1$.  The arguments  for this are  simple,
since  that time roughly  measures the interval  for the  jet shock to
cross a jet   radius.  Not until that   shock passes through  a  given
region of the jet can the jet react to the existence of the  external
medium.  Previous studies  of shocks interacting  with discreet dense clumps or
clouds (Klein, McKee \& Colella 1993; Jones \&  Kang 1993, Xu \& Stone
1995) and supersonic gas  ``bullets'' (Jones, Kang \& Tregillis  1994)
have  shown that the cloud  is  quickly
decelerated on this characteristic time scale  
(once the equivalent shock passes  through the cloud; 
see especially   Klein,    McKee  \& Colella  1993).
Precessing   jets  with  a sufficiently    large   angle, $\theta$, and
precession  rate, $\Omega$, will  behave  much like discrete clouds  in
this  respect, as noted earlier. There  are some constraints that must
be satisfied for this to make sense, however. As mentioned in \S 3 one
requires  $\sin{\theta} >   1/\sqrt{\chi}$ to  establish  that forward
portions of  the jet on opposite sides  of the precession  cone do not
overlap  at $t  \sim  1$. In  addition,  the  pitch, $\delta_z$ should
exceed   the    jet   diameter;  thus,   $\tau_p    >   2/(\sqrt{\chi}
\cos{\theta})$.  When those two conditions are satisfied we can simply
estimate  the time for the jet  shock to break out  of the rear of the
jet  by examining the flow  in  the frame of the   jet head. From that
perspective the flow seen  looking back to  the origin appears
as a series of gaseous disks, whose motion is sheared, due
to the precession. For small distances back from  the head, $\Delta x << v_o
\tau_p$, the transverse (shear) speed of  the  flow appears TO AN OBSERVER
MOVING WITH THE HEAD to be  $v_y
\approx  \Omega \Delta x \sin{\theta}$.   This transverse motion leads
to a ``thinning'' along the trailing edge of  the jet in the direction
back to the  source
and will allow the jet shock to break out of the rear of  the jet at a
displacement,   $\Delta y$,   after  $t  \sim  \sqrt{\Delta  y/(\Omega
\sin{\theta})}$.    If $\Delta y  = 1$  the jet shock will have
crossed a full radius of the jet.  This provides a time estimate,
$t_b \sim  1/\sqrt{\Omega \sin{\theta}}$, for the jet shock to
break out.  After this time we can consider the jet to have been
decelerated.  Note that the time $t_b$ is the same expression, except for
units, as the "disruption time" $t_d$ in Raga {\it et al.} 1992.
Applied to the case shown in  Figure 2, with $\theta =
26^o$ and   $\Omega = 27.3$,  we  would estimate  $t_b \sim  0.3$. The
earliest image in that  collage corresponds to  $t = 0.51$, and  it is
apparent    that the jet  shock   has  mostly penetrated the outermost
portion of the jet cross  section to the right  of the jet origin. The
leading jet material  on the left, which  was ejected  about 0.13 time
units later and,   thus, has an   age $\Delta  t  \approx 0.4$,  still
contains   the shock. But, it has    been substantially compressed. We
conclude, therefore, that our  simple formula reasonably  captures the
important features necessary to estimate the break out time.  Then, as
argued above, after $t_b$ we expect the jet  material to drop back and
eventually  collide  with  following    material, which   is   largely
un-decelerated. That can be seen clearly in Figure 2.  On the other hand
it is not fair to decribe the jet as disrupted.  So on that matter we do
not find support for the conclusion of Raga {\it et al.} 1992.

It is difficult to predict confidently from these simulations how
similar precessing jets will behave after much longer periods of time.
Clearly, the leading turns of the jet will begin to be swept up by
younger jet material. Perhaps that will eventually lead to something
resembling the conical jet simulated by Kochanek \& Hawley
(1990). But, it also seems very likely that as the forward-most
material in the jet moves ahead, some degree of disruption will begin
to take place. At least two effects suggest that. First, when the
leading turn of the jet is struck from behind, it tends to spread out,
as can be seen in both Figures 1 and 2.  In addition, Kelvin-Helmholtz
and Rayleigh-Taylor instabilities can eventually begin to influence
the jets, especially near the front.  The low density cavities that
form within the shroud, however, may reduce growth of disruptive
instabilities within the following jet material.

The global bow shock was identified by Cliffe {\it et al.} (1995) as
an important morphological feature. The bow shocks of individual
portions of the leading turn of the jet merge to create this
structure.  Its importance comes from the enlarged volume enveloped by
it and from the fact that flows within it, as we shall see, can be
distinctly different from those inside the simpler bow shocks of
straight jets.  One can imagine its formation as the outcome of the
penetration of the initial bow shock by younger portions of the first
turn of the jet.  That is illustrated well in Figure 4, which shows
the density and pressure cut through the $x-z$ plane for model 3.  It
is simple to make a rough estimate of the conditions for this to
occur, by asking how long it takes for those portions of the jet
ejected after half a period ($t = \tau_p/2$) to intersect the bow wave
of the material ejected at $t = 0$. They are on opposite sides of the
precession cone, of course. With the simplifying assumption that the
leading bow shock is conical with an opening half angle $\psi =
\arcsin{1/M_j}$ similar triangles can be constructed to show that the
younger material penetrates the leading bow shock after a time
\begin{equation}
{t \over \tau_p} \gsim {1\over 2} \left 
[\tan{(\theta + \phi)}\over{\tan{(theta)}}\right ].
\label{pentime}
\end{equation}
After this time it is reasonable to suppose that a global bow shock has formed.
For all of our precessing jets the time define by equation \ref{pentime}
is less than $\tau_p$, so that
penetration occurs within one precession period. One must also require
that the precession angle satisfy $\sin{\theta} > 1/\sqrt{\chi}$, as before,
which is slightly more restrictive in our cases.

\subsection{Kinematics \& Entrainment}

One of the key reasons for considering precessing jets as possible
drivers for molecular outflows is that the kinematics of gas swept
into the outflow by the jet may be expected to be very different from
that of straight jets, as we discussed earlier.
In fact, Chernin and Masson (1993) focused on the bow shock and 
its accompanying shroud
as the key dynamical elements in jet-driven molecular outflows.
In their numerical study of jet
driven molecular outflows Chernin {\it et al.} (1994) demonstrated that
a high Mach number, heavy radiative jet will accelerate material 
exclusively in the bow shock.  This kind of ``prompt entrainment'' must be
distinguished from ``steady state'' entrainment that occurs through
instabilities along the length of the jet beam (DeYoung 1986, 1993).  
As was noted in \S 2 such turbulent entrainment models have been invoked to 
explain molecular outflows.  In the results we present below we will also
focus on prompt entrainment of the ambient material. However, as we will
demonstrate, a new mode of entrainment also presents itself 
in precessing jets.  
The key features of the gas kinematics within the jet shroud are visible
in Figure 5, which is a slice in the $x-z$ plane for run 3, showing
flow velocity vectors on top of a gas density gray scale image. The 
pressure structure for the same jet is shown in Figure 4. 
Motions are rather complex, as might be expected.
Nonetheless, we can identify several distinct
kinematical elements that map onto the morphological features
discussed in the last section.

The material directly behind the global bow shock has a velocity structure 
similar to that seen in straight jets.  As the first turn
of the jet beam propagates
outward it accelerates the ISM through the bow shock.
That material is pushed aside as the jet continues propagating.  The 
shroud then accretes more ISM material 
as it penetrates outward.  Thus, we see larger forward-directed 
velocities near the top of the helix, where the
first turn of the beam interacts directly with the ISM, and
small transverse  velocities near the base of the shroud.

The high-density gas within the jet that has already passed through
the jet shock forms a second kinematical element.  Note that, like the
unshocked jet material, these shocked gas parcels also travel
in the radial direction into which they were originally
ejected.  It is compressed, and only begins to be significantly decelerated
once the jet shock ``breaks out'' of that jet segment, as discussed earlier.
There is also some expansion of the shocked jet, similar to that seen in
supersonic ``bullet'' calculations (Jones, Kang \& Tregillis 1994).

As the jet precesses, beam segments traveling 
radially outward along the 
precession cone create low-density, low-pressure wakes
directly behind them, with strong rarefaction waves extending to meet
the bow shocks of the segments in the next turn of the
helix.  These wakes were pointed out earlier and
can be seen in the Figures 2-4.  There is as much as a factor of 5
pressure change across them.  Such low
density zones are not seen in straight jets, because there is no
place for them geometrically.
The wakes in the precessing jet simulations
have the effect of ``entraining'' shocked ambient
material in the shroud to form another distinct
kinematic element. These regions, which we call the `wakes,'  
actually wrap around the jet material, like an
entwined corkscrew.  Because of the way these evacuated regions form
behind and follow the radially moving jet parcels, the jet's
precession gives the ambient material more forward directed motion
than is seen in a straight jet.  

Another kinematic element forms through the action of the secondary
bow shocks preceding younger turns in the jet. These also accelerate
material, effectively sweeping it along with the jet. Thus, we
see several ways in which the precessing jet,
as predicted by Chernin \& Masson (1994), 
deposits momentum with the ambient medium in more places than
at the leading head of the jet or through turbulence at the
boundaries.  

All of the simulations discussed so far were based on an equation of
state with $\gamma = 5/3$. As mentioned in \S 3 we have made an
initial attempt to understand how strong cooling would affect these
results by carrying out a simulation identical to run 3 except
that an ``isothermal'' (or nearly so) equation of state, $\gamma = 1.1$,
was used. The most obvious differences come through the fact that 
we expect much greater compression through the bow shock in this case
and a slower expansion of the bow shock laterally. These differences
are clearly evident in Figure 6, which represents the flow at approximately
the same time as shown in Figures 4 and 5. The global bow shock is
still present, but it ``hugs'' the jet much more closely. Regions of
high pressure are more limited to being nearly inside and in direct
contact with the leading turn of the jet. These differences have
a remarkable effect on the kinematics of shroud material, as will
become apparent in the following section.

A final element of the velocity structure of the gas seen in our
simulations is that of ``classically'' entrained ambient gas.  This is
shroud gas that mixed with beam material in the boundary of the jet.
It seems likely, even though large scale ($\lambda \sim 1$),
destructive instabilities take too long to form to influence these
flows on the timescales we have considered, that small scale mixing
will develop and some kind of a boundary layer will form. Numerically
this is unavoidable, because of numerical diffusion.  This numerically
entrained gas can be recognized in the flows as material with
densities intermediate between the jet and the ambient medium values
and having large velocity vectors pointed in the direction of the jet
flow. In the following section we will attempt to model the momentum
distribution predicted matter swept up by the precessing jets that we
have simulated. We cannot entirely eliminate the effects of numerical
diffusion in such a calculation, but we can try to understand its effects in two
ways. First, since numerical diffusion is reduced in higher resolution
simulations, we can compare results for the two different resolutions.
Inspection of the boundary layer between the jet and the ambient medium
shows the the higher resolution runs have a smaller transition width
between the jet and ambient medium relative to the jet radius.  Thus
less material is entrained in the high resolution simulations. The
thickness of this layer is only a few zones ($\sim 3-4$) across
in each case, so we expect little mass to be involved. As a
second test of this issue we have performed a 2-D cylindrical
calculation of a straight jet using the same TVD algorithm, including
a passively advected tracer that as set to unity in material injected
by the jet and set to zero for ambient matter. This allows us to
identify material that has kinematical properties of the jet but
originated in the ambient medium.  Plots of average momentum both with
and without this numerically mixed material are effectively
indistinguishable for those simulations.  We conclude, therefore, that
numerical diffusion effects are unimportant when considering momentum
distributions in our simulated flows.  We note however that numerical
mixing will be more severe in simulations where the flow moves oliquly across
the computation grid.  We had, unfortunately, not implemented a fluid tracer in 
the 3-D code when the simulations were run. We plan to include such a routine
in the next stage of this project which will include radiative cooling.

\section{Comparison with Observations}
In this section, we present a qualitative comparison between our
 precessing jet simulations and observations of molecular outflows.
 Since our calculations use simplifying assumptions (constant density
 ISM, polytropic flow), we do not expect a direct or quantitative
 comparison to be valid.  However, we will seek to compare and
 contrast the results that are found for precessing jets with those
 found for straight jets, and to comment on how some observational
 features may be better accounted for by precessing jets.

\subsection{Shape of Bow Shock}

Molecular outflows show a range in their degree of collimation with a
 wide variety of observed length-to-width ratios.  This
 characteristic is not easily reproduced by mechanisms involving
 straight jets, particularly straight jets with significant cooling,
 since their bow shocks/shrouds are not inflated around the beam.
 Thus the outflow ``lobes'' produced in these models are quite narrow.
 Wind-driven outflow mechanisms, on the other hand, have the opposite
 problem in that their lobes tend to be too wide to encompass those
 molecular outflows with lengths that are many times their widths.
 Our polytropic precessing jets clearly reproduce the wider lobes seen
 in some molecular outflows.  Since the width varies with cone angle,
 there is room in the model for a range of observed length to width
 ratios.  We see from our simulations that even in the case where
 $\gamma = 1.1$, crudely simulating the effects of radiative cooling,
 the bow shocks of the precessing jet appear quite wide although in
 this case the lobe is not inflated by high pressure inside the bow
 shock.  We see wide lobes in this case as a direct result of the
 precession and the global bow shock.  The length-to-width ratio of
 the widest point in the lobe depends on the cone angle of the jet.
 The wide variation in the molecular outflow geometries around YSOs
 can, therefore, be easily accounted for by assuming a range of jet
 precession cone angles.

In \S 2 we discussed the observational result that molecular
outflows present flow patterns with most of the material being
``forward driven''.  This means the mass moves primarily along the
direction defined by the the long axis of each lobe. Despite the fact
that the lobes in our simulations are wide, their momentum vectors are
primarily forward driven, in agreement with observations. To
demonstrate this effect in table 2 we present ``synthetic observations''
of the blueshift fraction ($f_{blue}$) of our simulated jet-driven
outflow lobes.  The quantity $f_{blue}$ is computed by first choosing
an inclination angle for the lobe relative to an observer's line of
sight.  The velocity vectors in the lobe are then projected onto the
line of sight and the blue-shifted and red-shifted components are then
mass-weighted and summed.  The fraction of blue-shifted material
gives a measure of the degree to which the lobe has either strong
transverse motions ($f_{blue} \sim .5$) as compared to primarily
forward driven motions ($f_{blue} \sim 1.$).  In table 2 we present
this fraction at two characteristic velocities: $V = 0.5V_{\rm
\scriptstyle max}$ and $V = V_{\rm \scriptstyle max}$ and two
inclination angles with respect to the line of sight: $30^{\rm o}$ and
$60^{\rm o}$. Here $V_{\rm \scriptstyle max}$ is the maximum projected 
velocity of the lobe.
We note that the while numbers presented here are indicative
of the trends seen in the simulations they are not intended as a
serious comparison with observations.  Such a comparison will have to
wait for more detailed modeling.

Table 2 compares the straight and precessing jet simulations for both
$\gamma = 5/3$ and $\gamma = 1.1$ runs. Consideration of the $\gamma = 5/3$
runs in table 2 demonstrates that there is a higher fraction of
blueshifted (forward driven) material in the lobes of a precessing jet
than in its straight counterpart.  For example, at an inclination
angle of $30^{\rm o}$ the $\gamma = 5/3$ straight jet has a blueshift
fraction of $61 \%$.  This means the contrast of blueshifted to
redshifted gas is only about a factor $1.5$. The precessing jet
simulated in run 1 however has a blueshift fraction of $96 \%$ which
represents 24 times more blueshifted material than redshifted
material.  Having so little gas moving away from the observer at this
inclination implies little lateral expansion of the lobe, which in turn
implies primarily forward driven motions.  This is in agreement with
the observations of Lada \& Fich (1995), who found the NGC 2264G
molecular outflow to have a significant blueshift to redshift contrast
(their Fig. 12).  One notable exception, however, occurs in the $V =
V_{\scriptstyle max}$ values in run 3.  This is the wide angle slow
precession run.  The low values at the maximum velocity are most
likely due to projection effects, where the highest velocity material
is sampled at opposite sides of the lobe, giving both red and blue
shifted components.  We note that Lada \& Fich's (1995) data showed a
similar downturn in the blueshifted fraction at the highest
velocities.

We note also that the $\gamma = 1.1$ straight jet has a larger
blueshift fraction than its $\gamma = 5/3$ counterpart.  This is
expected with the loss of thermal pressure
support behind the bow shock and indicates that even straight
radiative jets may be able to account for the forward driven
kinematics seen in molecular outflows.  The precessing jet still shows
marginally higher blueshift fractions, however, and with better
resolution and a more realistic treatment of cooling we should be able
to accurately discriminate between the two cases.

\subsection{Momentum Distribution}
In their study of the distribution of momentum in protostellar
molecular outflows Chernin \& Masson (1995) found that in most
outflows the peaks of mean momentum ($d{\bar P}/dz$) lie near the middle
of each lobe, with minima near the source (central star) and the heads
of the outflow.  Such a momentum distribution is difficult to
reproduce using a wide-angle wind model, a steady-state jet model or a
straight jet/bow shock model (see their Fig. 3 and 4).

With the density and velocity information from our simulations we
computed ``synthetic observations'' similar to those of Chernin and
Masson.  At each position along the z axis, we summed the total
momentum ($\rho v$) through the lobe for each $x$ and $y$.  Only
material with a density less than or equal to ${{(\gamma+1)} \over
{(\gamma-1)}}$ was included, since this represents the maximum density
of shocked ambient material in the strong shock limit.  This
characterization matches Chernin \& Masson's observations, which were
restricted to CO gas from the molecular cloud (i.e. excluding any jet
material).  We then found the width of the lobe on the sky at each
$z$-value.  Following Chernin \& Masson (1995), we divided the summed
momentum at each $z$ by the lobe width at that position to obtain a
value for the {\it total} mean momentum, $d{\bar P}/dz$, as a function of the
position along the lobe.  We also calculated the {\it
projected} mean momentum as a function of position.  In this case, rather
than using the total velocity in the momentum term, we used the
projected velocity; that is, either $v_{proj} = v_x$ or $v_{proj} =
v_y$, depending on the direction of the sight line.  This quantity
more closely represents Chernin \& Masson's (1995) actual
measurements, since they sampled only radial velocity components in
their radio beam.  They argue, however, that because most of the
momentum is forward directed, the projected momentum represents the
total momentum.  Figure 7 shows our calculations of
the distributions of total and projected momentum for all of our
$128^3$ resolution runs.
A comparison of the image pairs in Figure 7 (average
total ({\it left}) and projected ({\it right}) momentum), however,
reveals differences between these two quantities.  Probably the best
value to use, given Chernin \& Masson's (1995) argument as well as the
fact that our projected momenta are calculated for a viewing angle
perpendicular to $z-$axis, is a blend of the projected and total
momenta. The side-by-side display of both measures in Figure 7 should
give the reader a means to judge and compare the two individual quantities.

The precessing jet model does a better job of reproducing the
observations of Chernin and Masson than does the straight jet.  In the
straight jet the projected momentum of the straight jet peaks near the
end of the flow, with a sharp cutoff. The cutoff corresponds to the
bow shock, while the peak corresponds to the flow near the head of the
jet.  The momentum distributions of the precessing jet models,
however, have a cutoff that is less sharp (near the outside edge) and
are peaked closer to the center.  In some cases there are multiple
peaks. Each corresponds to shock structures attached to some portion
of the jet. The wider the cone angle, the more center-oriented the
momentum peak becomes. The best fit to the profiles presented by
Chernin \& Masson (1995) is run 3 with the wide cone angle and slow
precession rate.  The ability of precessing jets to model Chernin \& Masson's
observations is, most probably, accounted for by a combination of
effects that include the global bow shock and the presence
of material swept into the wakes
behind turns in the jet material.  Both of these
phenomena, unique to {\it precessing} jets, have the effect of
accelerating ambient material in the middle of the lobe. By contrast,
flows around straight jets are primarily accelerated at the head of
the jet and then swept around it.
 The momentum distribution associated with the precessing jet
with $\gamma = 1.1$ shows more exaggerated structure.  As in the
$\gamma = 5/3$ runs, the momentum distribution peaks closer to the
center of the lobe than for the corresponding straight jet.  Several
humps in the projected momentum distribution are seen along the
z-axis.  Each one of these humps corresponds to a point on the sky
where the jet beam is moving along the line of sight of the observer.  
These humps are
caused by the high velocity, higher density material in the shroud
wrapping around each turn of the beam.  The dominant hump corresponds
to the lead turn of jet material.  In our simulations, several humps
are seen because the jet has completed three half-periods of
revolution.  If a single half turn had been completed, there would be
one hump at the point along the axis where the bend in the beam of the
jet was oriented directly towards the observer, leading to a momentum
distribution much like that seen in the six objects studied by Chernin
\& Masson (1995) .

In the total momentum distributions a pronounced hump immediately
behind the end of the flow (Figure 7) always occurs.  Examining 2-D
cylindrical jet runs with a passively advected variable to identify
ambient material, we excluded the hypothesis that this effect was due
to numerically entrained gas at the sides of the jet.  Further
examination revealed that the momentum hump corresponded to shocked
ambient material directly behind the widest point in the jet head.
After being compressed in the strong bow shock at the jet head,
ambient gas is further compressed as it flows past the region of high
temperature shocked jet material above the jet shock.  Because it is
near the top of the bow shock, these gas parcels have a high
z-velocity components.  In addition, as this material is being pushed
aside by the shocked beam gas it acquires a sideways velocity
component as well.  Together, these components provide this material
with the highest total velocity vectors of all the shocked ambient
flow.  This effect, combined with its high density, gives the region
just to the side of the jet shock the greatest momentum.  This
momentum hump is not as pronounced in the straight jet with $\gamma =
1.1$, where the decreased pressure in the head of the jet means that
ambient material is not being as strongly compressed.

\section {Conclusion and Discussion}

We have presented fully 3-D simulations of precessing jets with a
range of cone angles and precession rates.  We compare the resulting
flows with those of straight jets and find several unique
morphological and kinematic features.  The bow shocks of precessing
jets are composed of two elements. There is the bow shock of the first
turn of jet material, which interacts directly with the ambient medium
and envelops the entire structure to form a ``global'' bow shock.  In
addition, there are secondary bow shocks of following turns of jet
material.  The secondary bow shocks propagate into the cavity created
by the global bow shock and interact with already shocked ambient gas.
The global bow shock, because it encompasses an ever-expanding
corkscrew of jet material, can be quite wide. Depending on the cone
angle and precession rate it can be quite asymmetric.  The ambient gas
swept into the flow has a complex kinematical structure. These include
ambient gas shocked by the global bow shock, ambient gas doubly
shocked by the global and secondary bow shocks, ambient gas swept into
wakes of jet turns and entrained (numerically or by turbulence) gas.
The kinematical structure is significantly more complex than that of
previously well-studied straight jets.  We also find significant
deceleration of the first turn of the jet beam.  This timescale is set
by the interval required for the terminal, or jet shock to break
through the lead turn of the jet. That time is determined by the
period required for the jet shock to cross radially a jet
segment. That depends on the full set of model parameters ($M_j, \chi,
\theta$, and $\tau_p$), because it is controlled by the jet shock
speed and the rate at which a segment of the precessing jet is thinned
by shear.  On longer timescales mergers between the leading turns and
a succession of following turns seems likely.  Our simulations did not
last long enough for the kinds of disruptive instabilities
traditionally associated with supersonic jets to develop, since they
require for our models an order of magnitude longer time than that
for the lead turn
to be decelerated. We also find no evidence for disruption of the jet
material on timescales derived by Raga, Canto\', \& Biro (1993) due to
a "sideways" interaction of the jet with the external environment. 
The leading protions of the jet are shocked and decelerated on the 
timescales  Raga {\it et al.} predict however,
as with supersonic clumps, disruption requres several times this interval.

We have found that precessing jets are successful in reproducing many
of the crucial observations characteristic of molecular outflows.  In
particular, the precession of the jet allows for wide lobes of swept
up or ``promptly'' entrained material while still maintaining a high
degree of forward-driven momentum. The momentum distributions are
significantly better matches to observations than those of straight
jets or wind-driven outflows.

The simulations we have performed have confirmed the notion that
precessing jets have a rich dynamical, kinematical and morphological
structure, which we have only begun to explore here.  This study
illustrates where future work would be most useful in understanding
these complex flows and in bridging the gap between models and
observations of molecular outflows.
For example, the inclusion of a realistic cooling function will be
an important next step.
Although such a realistic treatment was beyond the scope of
this initial study, our $\gamma = 1.1$ simulation indicates that
cooling affects the model in important ways, even improving its
agreement with some observations (such as the contrast of
blueshift:redshift in the lobe).

From the limited exploration of precession angle and precession
rate we can see that the precession angle largely determines the
width-to-length ratio of the lobe. Thus, observations of molecular outflows
(along with estimates of age and outflow velocities) would provide an
means to better define the range of precession cone angles found in
nature. The
possibility that momentum distribution peaks in flows with strong cooling
represent individual turns of the jet material provides motivation to
try to establish realistic precession rates and examine their fits to observations.
Finally, we note that these simulations examined the dynamics of
precessing jet flows over a very restricted range of dynamical 
timescales. It is not sufficiently clear how these jets and the
material swept up by them will evolve on much longer timescales, particularly 
after the oldest turns in the jets begin to merge with younger ones.

We wish to thank Charles Lada and Lawrence Chernin for enlighting
discussions of molecular outflow dynamics.  Dongsu Ryu was generous
in making his 3D TVD code available to us for this study.
We appreciate the contributions that Joseph Gaalaas made in helping
with the graphics.  This work was supported by the NSF through grant
AST-9318959 and by the University of Minnesota Supercomputer Institute.
AF recieved support from NASA grant HS-01070.01-94A
from the Space Telescope Science Institute, which is operated by
AURA Inc under NASA contract NASA-26555.

\begin {table}[par1]
\caption {Model Jet Parameters } \label{tab:counts1820}

\begin {center}
\begin {tabular} {llllllll} \hline
{run} &  {$\gamma$} & { $\chi$} & {M} & {$\theta_{\scriptstyle 0}$} & {$t_{cross}$} & {$\tau_p$} & {$N_a$}\\ \hline 
1 & {5/3} &{\space80}& 10 & $12^{\rm o}$ & {\space 1.15} & {\space ${1 \over 2} t_{cross}$} & {\space 10}\\
2 & {5/3} &{\space80}& 10 & $12^{\rm o}$ & {\space 1.15} & {\space ${1 \over 5} t_{cross }$} & {\space 10}\\
3 & {5/3} &{\space80}& 10 & $26^{\rm o}$ & {\space 1.25} & {\space ${1 \over 2} t_{cross}$} & {\space 10}\\
4 & {5/3} &{\space80}& 10 & $26^{\rm o}$ & {\space 1.25} & {\space ${1 \over 5} t_{cross }$} & {\space 10}\\
5 & {1.1} &{\space80}& 10 & $26^{\rm o}$ & {\space 1.25} & {\space ${1 \over 2} t_{cross}$} & {\space 10}\\
6 & {5/3} &{\space80}& 10 & $12^{\rm o}$ & {\space 1.15} & {\space ${1 \over 5} t_{cross}$} & {\space 20}\\
7 & {5/3} &{\space80}& 10 & $26^{\rm o}$ & {\space 1.25} & {\space ${1 \over 5} t_{cross}$} & {\space 20} \\
\end {tabular}
\end {center}
\end {table}

%\begin {center}
%\begin {tabular} {lllllll} \hline
%{run} &  {$\gamma$} & { $\chi$} & {M} & {$\theta_{\scriptstyle 0}$} & {$\tau_p$} \\ \hline\hline \\
%\end {tabular}
%\end {center}
%\end {table}

\begin {table}[blueshift]
\caption {Fraction Blueshift at projected velocities $0.5_{Vmax}$ and $V_{max}$.
Note the first five runs have $\gamma = 5/3$.  The final two runs have  
$\gamma = 1.1$}
\label{blueshift}
\begin {center}
\begin {tabular} {lllll} \hline
run number & $0.5V_{\rm max}(30^{\rm o})$ & $V_{max}(30^{\rm o})$ 
& $0.5V_{max}(60^{\rm o})$ & $V_{max}(60^{\rm o})$ \\ \hline
straight & 0.654 & 0.613 & 0.724 & 0.974 \\
run 1 	 & 0.670 & 0.968 & 0.744 & 0.989 \\
run 2 	 & 0.739 & 1.000 & 0.828 & 1.000 \\
run 3	 & 0.717 & 0.054 & 0.789 & 0.068 \\
run 4	 & 0.671 & 0.899 & 0.788 & 0.998 \\
straight & 0.736 & 0.910 & 0.833 & 0.999 \\
run 5	 & 0.757 & 1.000 & 0.855 & 1.000 \\
\end {tabular}
\end {center}
\end {table}

\clearpage
\begin{center}
{\bf FIGURE CAPTIONS}
\end{center}
\begin{description}

\item[Fig. ~1] {Volume rendering showing the shape of the jet and
bow shock for runs (clockwise from top left) runs 1, 3, 4 and 2. The "emissivity"
is $log_{10} \rho$ and the "opacity" selects density ranges that highlight
the jet and bow shock material.  The
times shown are 1.113, 1.208, 1.214, and 1.113, respectively.}

\item[Fig. ~2] {Log density renderings, as in fig 1, showing
thin slices in the x-z plane of run 7 at times t = 0.514,
0.855, 1.20, and 1.538, showing the effects of deceleration and
the merging of the first and second turns of jet material.}

\item[Fig. ~3] {Log pressure contours overlain on log density for a slice in  
the x-z plane of run 7 at time t = 1.20. The intervals are logarithmic,
each level is a factor of five
from the adjacent level.}

\item[Fig. ~4] {Log pressure contours overlain on log density for a slice in 
the x-z plane of run 3 at time t = 1.208. The intervals are logarithmic,
each level is a factor of five
from the adjacent level.}

\item[Fig. ~5] {x-z velocity vectors superposed on log density for a slice in
the x-z plane of run 3 at time t = 1.208. The dots represent the
tails of the arrows. The figure shows how ambient gas is swept up into
the evacuated cavities, or wakes, behind the radially ejected jet material.}

\item[Fig. ~6] {Log pressure contours overlain on log density for a slice in 
the x-z plane of run 5 at time t = 1.211. The intervals are logarithmic,
each level is a factor of five
from the adjacent level.}

\item[Fig. ~7] {Synthetic observations of total momentum (left column) and
projected momentum (right column) per unit length as a function of
position along the lobe for (from the top) a $\gamma = 5/3$ straight jet,
$\gamma = 1.1$ straight jet, precessing jets runs 1, 2, 3, 4, and
5. Momentum included is that of swept up ambient material only. The position
coordinates are given in zone numbers.}

\end{description}

\clearpage
\end{document}